\documentclass[12pt]{article}
\usepackage{amssymb}
\usepackage[dvips]{graphicx}
\title{Interpolating greedy and reluctant algorithms}  % Enter your title between curly braces
\author{ Pierluigi Contucci$^a$, Cristian Giardin\`a$^a$, Claudio Giberti$^b$,\\
         Francesco Unguendoli$^c$  and Cecilia Vernia$^c$ \\
        \small $^a$ Dipartimento di Matematica, Universit\`{a} di Bologna\\
        \small Piazza di Porta S.Donato 5, 40127 Bologna, Italy\\
        \small {\it \{contucci, giardina\}@dm.unibo.it}\\
        \small $^b$ Dipartimento di Informatica e Comunicazione, Universit\`a dell'Insubria,\\
        \small via Mazzini 5, 21100 Varese, Italy\\
        \small {\it claudio.giberti@uninsubria.it}\\
        \small $^c$ Dipartimento di Matematica Pura ed Applicata, Universit\`{a} di Modena\\
        \small e Reggio Emilia, via Campi 213/B, 41100 Modena, Italy\\
        \small {\it \{unguendoli, vernia\}@unimore.it}}
\date{}          % Enter your date or \today between curly braces
\begin{document}

\maketitle

\vskip .3cm
\begin{abstract}
\noindent
In a standard NP-complete optimization problem we introduce an interpolating
algorithm between the quick decrease along the gradient (greedy dynamics)
and a slow decrease close to the level curves (reluctant dynamics).
We find that for a fixed elapsed computer time the best performance of
the optimization is reached at a special value of the interpolation parameter,
considerably improving the results of the pure
cases greedy and reluctant.
\end{abstract}
%%%%%%%%%%%%%%%%%%%
\section{Introduction and results}
Combinatorial optimization stands as one of the most fruitful
fields in the intersection of applied and pure mathematics: it
connects the theory of the computational hardness to the
techniques widely used in the search for global minima for complex
functionals, i.e. functions with many local minima. Over the years
there have been many strategies proposed to solve efficiently hard
computational problems \cite{Fl,AMO,CCPS}. Among them the
statistical mechanics approach \cite{MPV,Ni} has opened new
interesting perspectives. In this paper we study the interpolation
between two algorithms, the greedy and the reluctant. The first is
the standard decrease along the deepest descent direction, while
the second is the closest decrease to the level lines. In previous
works \cite{Noi1,Noi2} we have studied and compared the two
algorithms focusing on relaxation time and minimum reached level.
We observed, moreover, how a simple convex interpolation between
them could improve the performances of both of them in large size
regime. In this work we push further such analysis introducing a
{\it smooth interpolation} between greedy and reluctant depending
on a parameter $\lambda$: small $\lambda$ plays the role of the
greedy algorithm, while large $\lambda$ that of reluctant. This
allows us a better tuning among the two and especially a parameter
optimization. The newly introduced algorithm is tested against a
model which has become the standard of NP complete problems: the
Sherrington Kirkpatrick model  of the mean field spin glass. Our
results confirm and extend those in \cite{Noi1,Noi2}: we find that
the relaxation time grows linearly when the algorithm is close to
the greedy regime and quadratically when it is close to the
reluctant one with a progressive condensation for large values of
$\lambda$. The dynamics is then tested in the search for low
energy configurations for fixed values of initial conditions,
where the reluctant dynamics works substantially better than any
other. The main result of this work is then the minimization at
fixed elapsed computer time. In this case, in fact, we find that
in the small size regime (compared to total search time) the
greedy component performs better than any other due the short
relaxation time and the fact that, basically, the dynamics is able
to find the ground state or, at least, to get very close to it.
Moreover and more interestingly, we find that increasing the
system size does not lead to a uniform deteriorating of the greedy
performance toward an improvement of the reluctant one. We find,
in fact, an optimal value of $\lambda \sim 10$, for which the
lowest energies are reached against a rather poor performance of
the greedy $(\lambda \sim 1)$ and reluctant algorithm $(\lambda
\sim 100)$. This optimal value appears to be independent of the
size.

\section{The model and the algorithm}
Let us consider the Sherrington-Kirkpatrick model \cite{SK}
defined by the
Hamiltonian
\begin{equation}
H(J,\sigma)=-\frac 12\sum_{i,j=1}^N J_{ij}\sigma_i\sigma_j
\end{equation}
where $\sigma_i=\pm 1$ for $i=1,\ldots ,N$ are spin variables
which interact through  an $N \times N$ symmetric matrix with
$J_{ij}$ independent, identically distributed gaussian random
couplings $(J_{ij}=J_{ji}, J_{ii}=0)$ with zero mean and variance
$1/N$. We focus our attention on a stochastic energy-decreasing
dynamics that, starting from any initial spin configuration at
time $t=0$ (which we choose at random with uniform distribution),
ends up on a local energy minimum. The evolution rule is:
\begin{enumerate}
\item Let $\sigma(t)=(\sigma_1(t),\ldots,\sigma_N(t))$ be the spin
configuration  at time $t$.
\item Calculate the spectrum of energy
change obtained by flipping the spin in position $i$, for
$i=1,\ldots,N$:
\begin{equation}
\Delta E_i=\sigma_i(t)\sum_{j\ne i} J_{ij}\sigma_j(t).
\end{equation}
If $\Delta E_i > 0$, $\forall i=1,\ldots,N$, then the algorithm
stops ($\sigma(t)$ is a local minimum).
\item Generate a random
number $D$ with probability density
\begin{equation}
f(x)=\left\{ \begin{array}{ll}
   \lambda e^{\lambda x} & \textrm{if } x\leq 0\\
   0  & \textrm{if }  x > 0
\end{array}\right. ,\qquad \lambda > 0.
\end{equation}
\item Select the site $i^{\star}$ associated with the closest energy
change to the value $D$, i.e.:
\begin{equation}
i^{\star}=\Bigg\{i\in\{1,\ldots,N\}:\Delta
E_{i^{\star}}=\min_{i\in\{1,\ldots,N\}} \{ |\Delta E_i-D|:\Delta
E_i<0\}\Bigg\}.
\end{equation}
\item Flip the spin on site $i^{\star}$:
\begin{equation}
\sigma_i(t+1)=\left\{ \begin{array}{ll}
  -\sigma_i(t) & \textrm{if } i=i^{\star}\\
   \sigma_i(t) & \textrm{if } i\neq i^{\star}.
\end{array}\right.
\end{equation}
\end{enumerate}
This algorithm generates a dynamics that, following a $1$-spin
flip decreasing energy trajectory, arrives at a $1$-spin flip
stable configuration, that is a configuration whose energy cannot
be decreased by a single spin-flip. The speed of convergence to
local energy minima is tuned by $\lambda$, the control parameter
in the probability distribution function for the move acceptance.
Of course, the larger is $\lambda$, the bigger it is the
probability of doing small energy-decreasing steps, so that the
trajectory will follow an evolution path close to level curves
(reluctant). On the other hand, small values of $\lambda$ enrich
the probability of large negative energy steps, which will quickly
drive the dynamic to the end-point (greedy). In the next Section,
by varying the control parameter $\lambda$, we study the
efficiency of the algorithm by measuring the average time to reach
a metastable configuration and the lowest energy value found as
the system size is increased.

\section{Results}

We performed a set of experiments for different values of
$N$, starting from $N$ initial conditions (for a system of size
$N$) and averaging the data on $nreal=1000$ disorder realizations.
We probed basically two quantities to measure the
performance of the algorithm:
\begin{itemize}
\item the average time (i.e. the number of spin flips)
to reach a minimum energy level
\begin{equation}
\tau=\frac 1M\sum_{i=1}^{M}t_i,
\end{equation}
with $M=N\cdot nreal$ and $t_i$, $i=1,\ldots,M$ the time
for each initial condition
\item the lowest energy found (averaged over disorder)
\begin{equation}
H_N=\left\langle\frac{\min_{\sigma}H_N(J,\sigma)}{N}\right\rangle_{nreal},
\end{equation}
where $\min_{\sigma}H_N(J,\sigma)$ is the minimum value of the
energy of the me\-ta\-sta\-ble states attained starting from $N$ initial
conditions.
\end{itemize}

\noindent
In Fig.1 we represent $\tau$ as a function of $N$
in the range $[25,300]$ for six distinct values
of $\lambda$ ($\lambda=1,10,25,45,70,100$),
together with the best numerical fits.
Because of high computational costs (which
increase with $\lambda$), the case
$N=300$ is studied in details only for $\lambda=1$ and
$\lambda=10$. On the other hand, the average time
has good self-averaging properties so that,
in order to have the trend for ``large'' $\lambda$
we focused on the case $N=300, \lambda=100$
with $nreal=140$ disorder realizations
instead of $nreal=1000$.
From Fig.1 we observe the progressive
increase of the slope in log-log scale from an almost linear law
in $N$ for $\lambda=1$ ($\diamond$) to an almost quadratic one for
$\lambda=100$ ($\ast$). Thus, this algorithm behaves as a
``smooth interpolation'' between the two deterministic
dynamics: ``greedy'', that we obtain for $\lambda=1$,
and ``reluctant'', here represented by $\lambda=100$.
More in detail, the numerical fits of Fig.1 are:
 $\tau_{\lambda=1}(N)\sim N^{1.027}$ ($\diamond$),
$\tau_{\lambda=10}(N)\sim N^{1.263}$ ($+$),
$\tau_{\lambda=25}(N)\sim N^{1.600}$ ($\square$),
$\tau_{\lambda=45}(N)\sim N^{1.796}$ ($\times$),
$\tau_{\lambda=70}(N)\sim N^{1.911}$ ($\triangle$),
$\tau_{\lambda=100}(N)\sim N^{1.932}$ ($\ast$). The fits are quite
good for all the cases but $\lambda=10$ and $\lambda=25$. In these
cases the quality of the fit is enhanced excluding the data
corresponding to $N=25$ and $N=40$. So we obtain
$\tau_{\lambda=10}(N)\sim N^{1.184}$ and $\tau_{\lambda=25}(N)\sim
N^{1.488}$.

\begin{figure}
    \setlength{\unitlength}{1cm}
          \centering
               \includegraphics[width=13cm,height=10cm]{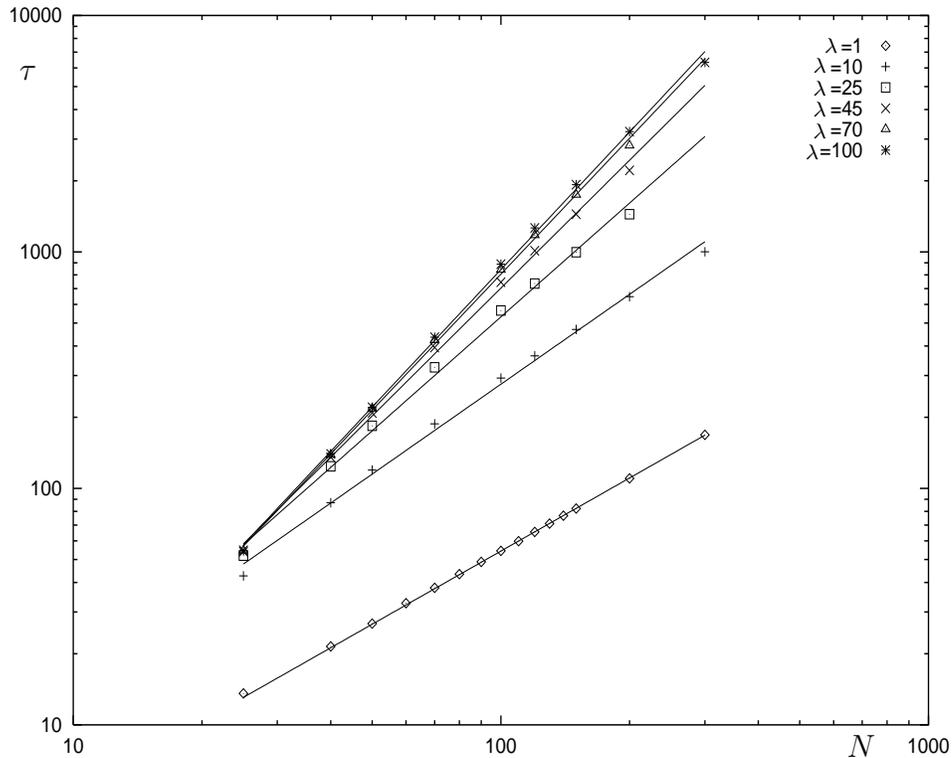}
               \put(-1.5,0){$N$}
               \put(-12.5,9){$\tau$}
               \put(-1.8,9.36){{\scriptsize{$\lambda$}}}
               \put(-1.95,9.1){{\scriptsize{$\lambda$}}}
               \put(-1.95,8.79){{\scriptsize{$\lambda$}}}
               \put(-1.95,8.51){{\scriptsize{$\lambda$}}}
               \put(-1.95,8.22){{\scriptsize{$\lambda$}}}
               \put(-2.05,7.95){{\scriptsize{$\lambda$}}}
               \caption{Average time $\tau$ to reach a metastable
               configuration as a function of $N$ for
 $\lambda=1$ ($\diamond$), $\lambda=10$ ($+$), $\lambda=25$
($\square$), $\lambda=45$ ($\times$), $\lambda=70$ ($\triangle$),
$\lambda=100$ ($\ast$).}
\end{figure}

\newpage
Next, we measured the lowest energy found by the algorithm. In
Fig.2 we represent $H_N$ as a function of $N$ for different values
of $\lambda$ and for a fixed number of $N$ initial conditions.
While, for small $N$, the ground state is believed to be closely
approximated for all values of $\lambda$ (in fact, varying
$\lambda$, the lowest energy $H_N$ undergoes a relative change of
$2.5\cdot 10^{-4}$ for $N=25$ and of $8.8\cdot 10^{-4}$ for $N=40$
and $50$), the best results for large $N$ $(>50)$ are obtained for
$\lambda=100$ which corresponds to deterministic reluctant
dynamics. Therefore, this confirms that, for a fixed number of
initial spin configurations, the algorithm that makes moves
corresponding to the ``smallest'' possible energy decrease is the
most efficient in reaching low-energy states. In other words, the
slower $\ldots$ the better! However, this is reflected in an
increasing cost for the computational time.

\begin{figure}
    \setlength{\unitlength}{1cm}
         \centering
               \includegraphics[width=13cm,height=10cm]{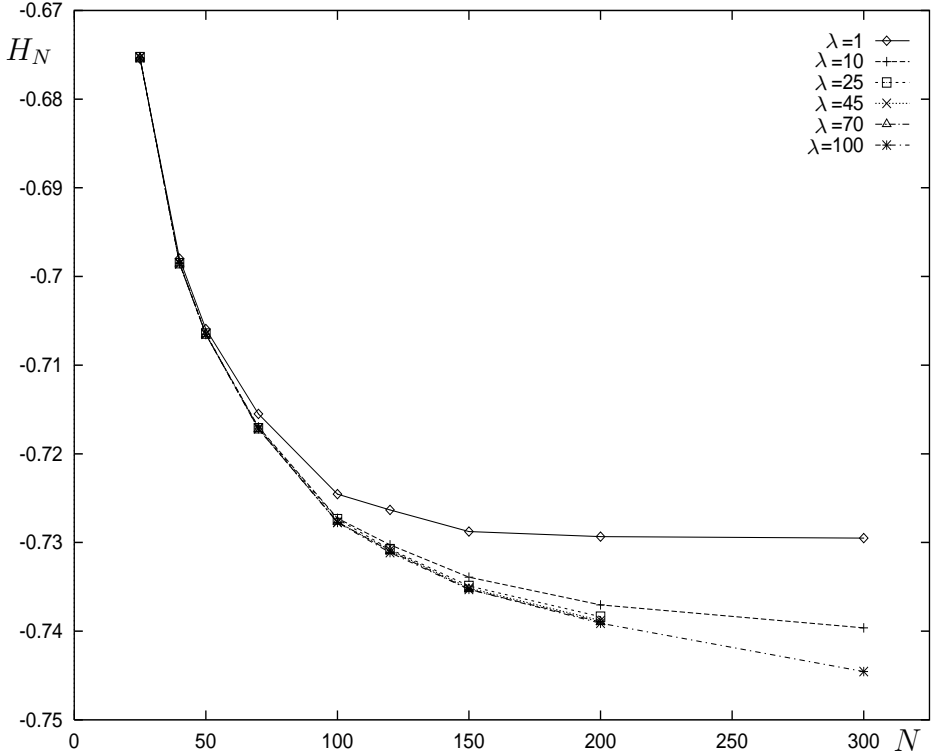}
               \put(-0.9,0){$N$}
               \put(-12.7,9.2){$H_N$}
               \put(-1.8,9.36){{\scriptsize{$\lambda$}}}
               \put(-1.95,9.1){{\scriptsize{$\lambda$}}}
               \put(-1.95,8.79){{\scriptsize{$\lambda$}}}
               \put(-1.95,8.51){{\scriptsize{$\lambda$}}}
               \put(-1.95,8.22){{\scriptsize{$\lambda$}}}
               \put(-2.05,7.95){{\scriptsize{$\lambda$}}}
               \caption{Lowest energy value $H_N$ as a function of $N$ obtained using
               a protocol with $N$  initial conditions for $1000$ disorder realizations
                for $\lambda=1$ ($\diamond$),
                $\lambda=10$ ($+$), $\lambda=25$ ($\square$), $\lambda=45$
($\times$), $\lambda=70$ ($\triangle$), $\lambda=100$ ($\ast$).}
\end{figure}

Indeed, when the analysis is focused on the performances for a
fixed elapsed time, the situation changes drastically. In Fig. 3
we compare the minimum energy values $H_N$, obtained considering
different system sizes and, for each of them, five different
parameter values ($\lambda=1, \lambda=10, \lambda=25, \lambda=45,
\lambda=100$) for an elapsed time of $50$ h of CPU on a IBM SP4.
The system size $N=350$ is studied only for $\lambda=10$,
$\lambda=25$ and $\lambda=45$. Each run (i.e. for fixed $N$ and
$\lambda$) consists of $1000$ disorder realizations, with the same
CPU time length ($3$ min.) assigned to each sample. For $N\le 100$
we believe to find the ground state of the system, since varying
$\lambda$ the values of $H_N$ coincide, within our numerical accuracy
($10^{-10}$).
The best result is obtained for the
case $\lambda=10$, which seems to be the best compromise to obtain
a dynamical trajectory that is able to arrive deep enough with
respect to energy levels but without wasting all the time in the
search for the slower of possible path. We note that this finding
is in good agreement with the result of previous analysis
\cite{Noi1}, where a convex linear combination of reluctant (with
probability P) and greedy (with probability $1-P$) dynamics was
considered. The optimal value $\lambda \sim 10$ is the one for
which the relaxation time $\tau \sim N^{\alpha}$ grows with a
scaling exponent $\alpha$ which is the closest, among the others,
to the value $\alpha=1.26$ of the optimal convex combination with
$P=0.1$.

\begin{figure}
    \setlength{\unitlength}{1cm}
          \centering
               \includegraphics[width=13cm,height=10cm]{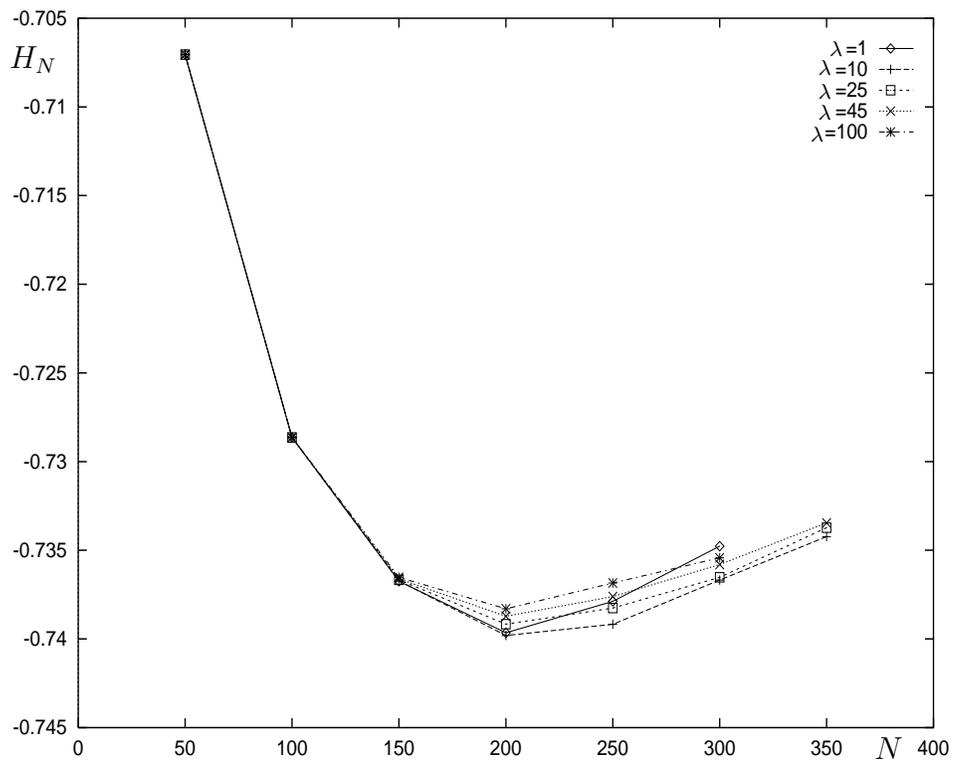}
               \put(-1.2,0){$N$}
               \put(-12.7,9.2){$H_N$}
               \put(-1.8,9.36){{\scriptsize{$\lambda$}}}
               \put(-1.95,9.1){{\scriptsize{$\lambda$}}}
               \put(-1.95,8.79){{\scriptsize{$\lambda$}}}
               \put(-1.95,8.51){{\scriptsize{$\lambda$}}}
               \put(-2.05,8.22){{\scriptsize{$\lambda$}}}
               \caption{Lowest energy value $H_N$ as a function of $N$
for a fixed CPU time of $50$ h on a IBM SP4
for $\lambda=1$
($\diamond$), $\lambda=10$ ($+$), $\lambda=25$ ($\square$),
$\lambda=45$ ($\times$), $\lambda=100$ ($\ast$).}
\end{figure}

Improvements of the greedy and reluctant algorithms
is presently under study \cite{Noi3}, by permitting also {\em
increase} in energy with exponential decrease in time, in the
very same spirit of the well-known Simulated Annealing strategies.

\vspace{0.3cm}
\noindent
{\bf Acknowledgments: } We thank the Cineca staff for the
technical support, Prof. S. Graffi and Prof. I. Galligani
for their encouragement toward our work. One of us (P.C.) thanks
the organizers of the Scientific Meeting OPT2003: ``Numerical Methods 
for Local and Global Optimization: Sequential and Parallel 
Algorithms'' for the invitation and the stimulating atmosphere.

\end{document}